\documentstyle[psfig]{mn}

\def\grb{gamma-ray burst}

\def\theta{\vartheta}
\def\phi{\varphi}
\def\sax{{\it Beppo}SAX}

\def\ltsima{$\; \buildrel < \over \sim \;$}
\def\lsim{\lower.5ex\hbox{\ltsima}}
\def\gtsima{$\; \buildrel > \over \sim \;$}
\def\gsim{\lower.5ex\hbox{\gtsima}}

\begin{document}

\title[Absorption in GRB afterglows]{On the role of 
extinction in failed gamma-ray burst optical/IR afterglows}

\author[Lazzati et al.]{Davide Lazzati$^{1,2}$, Stefano Covino$^2$
and Gabriele Ghisellini$^2$ \\
$^1$ Institute of Astronomy, University of Cambridge,
Madingley Road, Cambridge 
CB3 OHA, UK; e-mail: lazzati@ast.cam.ac.uk \\
$^2$ Osservatorio Astronomico di Brera, Via E. Bianchi 46, I-23807 
Merate, Italy; e-mail: covino,gabriele@merate.mi.astro.it }
\maketitle

\begin{abstract}
While all but one Gamma-Ray Bursts observed in the X-ray band showed
an X-ray afterglow, about 60 per cent of them have not been detected
in the optical band. We demonstrate that in many cases this is not due
to adverse observing conditions, or delay in performing the
observations.  We also show that the optically non-detected afterglows
are not affected by particularly large Galactic absorbing columns,
since its distribution is similar for both the detected and
non-detected burst subclasses.  We then investigate the hypothesis
that the failure of detecting the optical afterglow is due to
absorption at the source location.  We find that this is a marginally
viable interpretation, but only if the X-ray burst and afterglow
emission and the possible optical/UV flash do not destroy the dust
responsible for absorption in the optical band.  If dust is
efficiently destroyed, we are led to conclude that bursts with no
detected optical afterglow are intrinsically different.  Prompt
infrared observations are the key to solve this issue.
\end{abstract}


\section{Introduction}

The standard external shock synchrotron model (Meszaros \& Rees 1997;
Sari et al. 1998) has been very successful in describing the
properties of observed optical afterglows (Wijers et al. 1997; Galama
et al. 1998d, Covino et al. 1999).  However, for more than half of the
afterglows observed in the optical band we did not detect any
emission.  We will call these Failed Optical Afterglow (FOA) gamma-ray
bursts.  In all the \grb~error boxes promptly followed by narrow field
X-ray instruments, an X-ray transient (afterglow) has been detected
(with the only exception of GRB~990217), while only for $\sim 40$ per
cent of them optical observations have revealed an afterglow at
optical wavelengths.  This despite the rough similarity of the X-ray
afterglow fluxes and the prompt reaction of optical telescopes.
Paczynski (1998) ascribes this failed detection to dust extinction
pointing out how this interpretation requires the association of
bursts with star forming regions.  If this is the case, infrared
observations should be better suited for the hunt of afterglows, where
the extinction plays a reduced role.  For this reason, the IR
follow-up of GRBs has recently become quite common, and some
afterglows (GRB~990705, Masetti et al. 2000; and GRB~000418, Klose et
al. 2000a) have been detected in the infrared before being confirmed
at optical wavelengths.  Yet, we still miss the detection of an IR
afterglow without an optical counterpart: such a detection would
confirm the role of dust in FOAs.  Adding confusion to this picture,
observations of extinction in X-ray spectra seem to reveal a very low
gas to dust ratio (Vreeswijk et al. 1999, Galama \& Wijers 2000)
which, if common in all GRB environments, would strongly limit the
role of dust extinction in the absorption of afterglows in the optical
band and, even more, in the NIR.

In this paper we show that upper limits derived for FOAs are indeed
not consistent with an ``average afterglow'', contrary to what
recently claimed by Galama \& Wijers (2000).  We then analyze the
properties of detected and non detected optical afterglows in order to
check whether the absorption commonly seen in star forming regions can
explain the large fraction of FOAs.


\begin{table*}
\begin{tabular}{lllllllllll}
\hline
GRB  &$F_{X,\rm NFI}$  &$\Delta t_x$ &$\delta_x$ &Ref  &$R$   &$\Delta t_R$ &$\delta_R$ &Ref &$z$ &Ref \\
     &$10^{-13}$cgs  & h           &           &       &      &h            &           &    &    &      \\
\hline \hline 
970228.12362 &28$\pm$4    &8       &1.32       &Co97 &21.5$\pm$0.3   &16.5  &1.73$\pm$0.12 &Ma98,Ga00 &0.695   &Bl98  \\
970508.904   &7$\pm$0.7   &6       &1.1        &Am98 &19.77$\pm$0.1   &52   &1.2$^a$       &Pe98,Ga98a &0.835  &Me97   \\
971214.97272 &4$\pm$0.4   &6.7     &0.9        &An97 &22.06$\pm$0.06 &13    &1.20$\pm$0.02 &Di98      &3.418   &Ku98   \\
980326.88812 &NP          &---     &---        &---  &21.25$\pm$0.03 &11    &2             &Bl99      &---     &---        \\
980329.1559  &7.8$\pm$0.9 &7       &1.35    &Za98a,b &21.2$\pm$0.3$^b$ &17  &1.3$\pm$0.1   &Re99      &---     &---    \\
980425.90915$^c$ &4$\pm$0.6 &10    &0.2       &Pi00a &15.7$\pm$0.1   &59.8  &---           &Ga98b     &0.0085  &Ti98    \\
980519.51403 &1.4$\pm$0.3 &9.7     &1.8        &Ni99 &20.4$\pm$0.1   &15.5  &2.05$\pm$0.07 &Ha99      &        &   \\
980613.20215 &1.1$\pm$0.3 &9       &0.8        &Co99a &22.9$\pm$0.2   &16.3  &1             &Hj98,Dj98a &1.0964 &Dj98b    \\
990123.40780 &110         &5.8     &1.35  &He99a &18.26$\pm$0.04$^d$ &3.8   &1.12$\pm$0.03 &Od99,Ga99 &1.6004  &Ku99    \\
990510.36743 &14.7$\pm$1.8 &8      &1.4        &Ku00 &17.54$\pm$0.02 &3.5   &0.82$\pm$0.02 &Hr99      &1.619   &Vr99a   \\
990705.66765 &1.9         &11      &1.6        &Am00 &18.7$\pm$0.05$^e$ &5.5 &1.68$\pm$0.10 &Ma00    &---     &---   \\
990712.69655 &NP          &---     &---        &---  &19.4$\pm$0.1   &4.16  &0.97$\pm$0.02 &Sa00a     &0.4331  &Vr00 \\
001011.66308 &NP          &---     &---        &---  &20.6$\pm$0.1   &8.4   &1.4           &Go00      &---     &---    \\
\hline
980703.182468 &7.5        &22    &1.3$\pm$0.25 &Ga98c &21.00$\pm$0.09 &22.6 &1.39$\pm$0.3  &Ca99a      &0.9662  &Dj98c \\
990308.21883  &---        &---     &---        &---  &18.14$\pm$0.05 &3.34  &1.2$\pm$0.1   &Sc99      &---     &---  \\
991208.192269 &---        &---     &---        &---  &18.7$\pm$0.1   &49.9  &2.15          &Je99a      &0.7055  &Di99 \\
991216.671544 &1240$\pm$40 &4.03   &1.64       &Ta99 &18.49$\pm$0.05 &10.8  &1.22$\pm$0.04$^f$ &Ha00  &1.02    &Dj99 \\
000131.62446  &---        &---     &---        &---  &23.26$\pm$0.04 &84.3  &2.25$\pm$0.19 &An00      &4.50    &An00 \\
000301.41084  &---        &---     &---        &---  &20.42$\pm$0.06 &36.5  &1.18$\pm$0.14 &Sa00b     &2.0335  &Ca00a \\
000418.41921  &---        &---     &---        &---  &21.63$\pm$0.04 &59.3  &0.86$\pm$0.06 &Kl00      &1.1854  &Bl00 \\
000630.02145  &---        &---     &---        &---  &23.04$\pm$0.08 &21.6  &1.1$\pm$0.3   &Je00      &---     &---  \\
000911.30237  &---        &---     &---        &---  &20.26$\pm$0.17 &34.3  &1.5$\pm$0.14  &Pr00a,La00 &---     &---  \\
000926.99274  &2.1$\pm$0.6 &54.2   &4.3$\pm$1.0 &Pi00b &19.37$\pm$0.02 &20.7 &1.36$\pm$0.11 &Sa00c,Fy00a &2.066 &Fy00b\\ 
001007.20749  & ---       &---     &---        &---  &20.3           &83    &0$^g$         &Ca00b,Pr00b &---    &--- \\  
\hline
\hline
\end{tabular}
\begin{flushleft}
Notes: $\delta_\nu$ is defined by $F_\nu (t) \propto t^{-\delta_\nu}$. 
X-ray fluxes in the 2--10 keV band. NP=repointing of \sax~not possible.
$^a$: for $t>2$ days; earliest detection at 3.1 hours: $R=21.1\pm0.1$.
$^b$: $R$ mag derived from $I=20.8\pm0.3$; $R$=23.6$\pm$0.2 after 20 hours.
$^c$: = SN 1998bw, not used in the analysis.
$^d$: Converted from Gunn $r$-mag.
$^e$: $R$ mag derived from $H=16.57\pm0.05$.
$^f$: for t$\le$1.2 days.
$^g$: for t$\le$3.5 days, $\delta_R\sim1.4$ after.\\
Am98: Amati et al., 1998;
Am00: Amati et al., 2000
An97: Antonelli et al., 1997;
An00: Andersen et al., 2000;
Bl98: Bloom et al., 1998;
Bl99: Bloom et al., 1999;
Bl00: Bloom et al., 2000;
Ca99a: Castro-Tirado et al., 1999a;
Ca00a: Castro et al., 2000a;
Ca00b: Castro et al., 2000b;
Co97: Costa et al., 1997;
Co99a: Costa et al., 1999a;
Di98: Diercks et al., 1998;
Dj98a: Djorgovski et al., 1998a;
Dj98b: Djorgovski et al., 1998b;
Dj98c: Djorgovski et al., 1998c;
Dj99: Djorgovski et al., 1999;
Do99: Dodonov et al., 1999;
Fy00a: Fynbo et al., 2000a;
Fy00b: Fynbo et al., 2000b;
Ga98a: Galama et al., 1998a;
Ga98b: Galama et al., 1998b;
Ga98c: Galama et al., 1998c;
Ga99: Galama et al., 1999;
Ga00: Galama et al., 2000;
Go00: Gorosabel et al., 2000;
Ha99: Halpern et al., 1999;
He99a: Heise et al., 1999;
Hj98: Hjorth et al., 1998;
Hr99: Harrison et al., 1999; 
Je99a: Jensen et al., 1999a;
Je00: Jensen et al., 2000;
Kl00: Klose et al., 2000a;
Ku98: Kulkarni et al., 1998;
Ku99: Kulkarni et al., 1999;
Ku00: Kuulkers et al., 2000; 
La00: Lazzati et al., 2000;
Ma98: Masetti et al., 1998;
Ma00: Masetti et al., 2000;
Me97: Metzger et al., 1997;
Ni99a: Nicastro et al., 1999a;
Od99: Odewahn et al., 1999;
Pe98: Pedersen et al., 1998;
Pi00a: Pian et al., 2000;
Pi00b: Piro et al., 2000;
Pr00a: Price et al., 2000a;
Pr00b: Price et al., 2000b;
Re99: Reichart et al., 1999;
Sa00a: Sahu et al., 2000;
Sa00b: Sagar et al., 2000;
Sa00c: Sagar et al., 2001;
Sc99: Schaefer et al., 1999;
Ta99: Takeshima et al., 1999;
Ti98: Tinney et al., 1998;
Vr99a: Vreeswijk et al., 1999a;
Vr00: Vreeswijk et al., 2001;
Za98a: In't Zand et al., 1998a;
Za98b: In't Zand et al., 1998b;
\end{flushleft}
\caption{{Properties of the bursts with associated optical transient. 
The first 13 bursts have been observed by the \sax-GRBM/WFC, 
while the remaining bursts (below the horizontal line) have been 
discovered by other instruments (see text).}
\label{tab:uno}}
\end{table*}



\begin{table*}
\begin{tabular}{llllllll}
\hline
GRB &$F_{X,\rm NFI}$ &$\Delta t_x$ &$\delta_x$ &Ref &$R$ &$\Delta t_R$ &Ref  \\
    &$10^{-13}$cgs    & h          &           &    &    &h            &     \\
\hline \hline 
970402.930     &2.2$\pm$0.6 &8   &1.6  &Ni98  &21    &18.5  &Gr97a \\  
971227.34938   &2.6$\pm$0.6 &14  &1.12 &An99a  &22.8  &21.3  &Gr97b \\ 
981226.40793   &5$\pm$1     &11  &1.3  &Fr00  &23    &10.   &Li99  \\  
990217.22462   &$<1$     &6   &$>$1.6  &Pi99a &23.5  &19.   &Pa99a  \\ 
990627.20894   &3.5         &8   &---  &Ni99b &21.   &23.   &Ro99  \\  
990704.7294    &4.4$\pm$0.3 &8   &---  &Fe99  &22.5  &4.6   &Je99b  \\ 
990806.60286   &5.5$\pm$1.5 &7.8 &---  &Fr99  &22.   &3.8   &Vr99c \\  
990907.7319    &15$\pm$5    &11  &---  &Pi99b &22.9$^a$ &24.9 &Pa99b \\
990908.00125   &NP          &--- &---  &---   &20.$^b$ &11.5 &Ax99   \\ 
991014.9115    &3.5$\pm$0.5 &13 &$>$0.4 &Za00 &22.6  &12.9  &Ug99   \\  
991105.69495   &NP          &--- &---  &---   &23.5  &16.   &Pa99c   \\
991106.4545   &1.25$\pm$0.3 &8   &---  &An99b &21.   &9.1   &Ca99b   \\
000210.36396   &4.5         &7.2 &---  &Co99b &23.3  &16.   &Go00a,b   \\ 
000214.042    &2.75$\pm$0.9 &12  &0.6  &An00  &21.$^c$ &32.4 &Rh00 \\  
000424.76258   &---         &--- &---  &---   &22.8  &33.   &Ug00   \\  
000528.36568   &1.7$\pm$0.3 &8.3 &1    &Ku00  &23.3  &18    &Pa00a  \\  
000529.3361    &2.8$\pm$0.7 &7.5 &---  &Fe00  &22.3  &47    &Pa00b   \\ 
000615.2625    &---         &10  &---  &BS00  &21.5  &4.2   &St00   \\  
000620.2317    &---         &--- &---  &---   &19.8  &5.7   &Go00c   \\ 
\hline
990520.08539   &---         &--- &---  &---   &21.7$^d$&19.5  &Ma99  \\
991217.17496   &---         &--- &---  &---   &22.   &11.   &Mo99   \\ 
000416.6062    &---         &--- &---  &---   &20.7  &50.3  &Pr00c   \\ 
\hline
\hline
\end{tabular}
\\
\begin{flushleft}
Notes:NP=Repointing of \sax~not possible;
$^a$: $R$ mag derived from V$>23.2$.
$^b$: $R$ mag derived from $V>20.3$.
$^c$: $R$ mag derived from $K>18.15$.
$^d$: $R$ mag derived from $V>22$. \\
An99a: Antonelli et al., 1999a; 
An99b: Antonelli et al., 1999b;
An00: Antonelli et al., 2000; 
Ax99: Axwlrod et al., 1999; 
BS00: \sax~mail \# 00/18 = GCN Circ. \# 707; 
Ca99b: Castro-Tirado et al., 1999b; 
Co99b: Costa et al., 1999b; 
Fe99: Feroci et al., 1999; 
Fr99: Frontera et al., 1999;
Fr00: Frontera et al., 2000;
Go00a: Gorosabel et al., 2000a; 
Go00b: Gorosabel et al., 2000b;
Go00c: Gorosabel et al., 2000c;
Gr97a: Groot et al., 1997a;
Gr97b: Groot et al., 1997b;
Li99: Lindgren et al., 1999;
Je99b: Jensen et al., 1999b; 
Ma99: Masetti et al., 1999; 
Mo99: Mohan et al., 1999; 
Ni98: Nicastro et al., 1998;
Ni99b: Nicastro et al., 1998;
Pa99a: Palazzi et al., 1999a; 
Pa99b: Palazzi et al., 1999b; 
Pa99c: Palazzi et al., 1999c; 
Pa00a: Palazzi et al., 2000a; 
Pa00b: Palazzi et al., 2000b; 
Pi99a: Piro et al., 1999a;
Pi99b: Piro et al., 1999b;
Pr00c: Price et al., 2000; 
Rh00: Rhoads eta l., 2000; 
Ro99: Rol et al., 1999; 
St00: Stanek et al., 2000; 
Ug99: Uglesich et al., 1999; 
Ug00: Uglesich et al., 2000; 
Vr99c: Vreeswijk et al., 1999c; 
Za00: in't Zand et al., 2000.
\end{flushleft}
\caption{{Properties of the bursts with \sax-WFC detection but
without associated optical transient. The last three bursts below the 
horizontal line refer to the $\gamma$-ray poor GRBs (or X-ray transients) 
detected by \sax.}
\label{tab:due}}
\end{table*}


\section{Observations}
\label{sec:obs}

The first problem we face if we want to quantitatively describe the
failed detection of afterglows is the extreme inhomogeneity of the
sample.  Different burst error boxes have been observed with different
telescopes, with different depths, in different filters and at
different times after the burst explosion.  This situation makes
extremely difficult even to understand whether FOAs are consistent
with the brightness dispersion of the detected afterglows.  We
therefore attack the problem in two steps: first we ask whether the
upper limits and reaction times in the case of FOAs are really
inconsistent with those observed in optically detected afterglows,
then we analyze the X-ray properties of both optical detections and
non detections, to see if there is any difference.

Table~\ref{tab:uno} and Table~\ref{tab:due} report the data of the
bursts with and without a detected optical afterglow, respectively.
We considered all bursts with an optically detected afterglow,
irrespective if their locations has been provided by \sax~(first 13
bursts in Table~\ref{tab:uno}) or the IPN network and/or the XTE
satellite (remaining bursts in Table~\ref{tab:uno}). Instead, for
FOAs, we have been more restrictive and have considered only those
bursts detected by the WFC of \sax, which usually gives narrower error
boxes, facilitating the search for an optical transient in the field
of view. It is not appropriate to include the IPN-detected GRBs in the
sample of FOAs, since some of the error boxes have been observed only
partially, so that it is not possible to define a single limiting
magnitude for each burst. It is possible, even though not required by
the data (see the following statistical analysis), that the subsample
of IPN OAs is intrinsically different from the sample of \sax~OAs. For
this reason, we will compare in the following the FOAs both with the
sample of all OAs and with the subsample of \sax~OAs.  The results
will be similar, but the larger sample of OAs will give higher
statistical confidence, due to the larger number of elements.

\begin{figure}
\centerline{\psfig{file=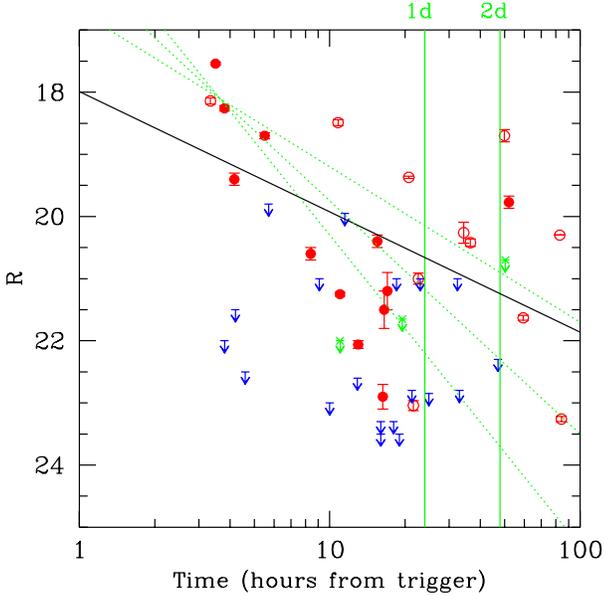,width=8cm}}
\caption{{Detection $R$ magnitude (or upper limits) versus the time 
of observation for a set of afterglows.  Filled circles show optical
detections of \sax~afterglows while empty circles show detections of
non \sax~afterglows.  Arrows show upper limits for \sax~failed optical
afterglows.  Arrows with crosses refer to the upper limits on
$\gamma$-ray poor X-ray transients detected by \sax.  The dark solid
line is the best fit for the magnitudes of detections vs. time.
Dotted lines show the $F_\nu(t)\propto t^{-1}$, $t^{-1.5}$ and
$t^{-2}$ relations.}
\label{fig:opt}}
\end{figure}

These data have been used to produce Figure~\ref{fig:opt}, which shows
the magnitudes of the optical afterglow detections and upper limits,
all in the $R$ band\footnote{For those bursts that do not have any $R$
band measurement, a spectrum $F(\nu)\propto\nu^{-1}$ has been assumed
to transform $V$, $I$ and $H$ magnitudes in $R$ magnitudes.}  versus
the time of observation.  Filled and empty circles correspond to
\sax~and non-\sax~bursts with detected optical afterglows, while
arrows are upper limits.  The three X-ray transients tentatively
associated with bursts ($\gamma$-ray poor GRBs, non significantly
detected in the GRBM) are included in the sample of upper limits
(however, their exclusion does not significantly influence any of the
following results).

\begin{figure}
\centerline{\psfig{file=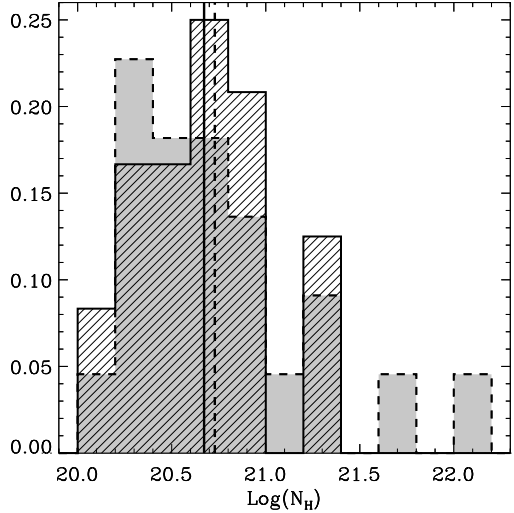,width=8cm}}
\caption{{Logarithmic distribution of the column densities of bursts with
(solid line and line shaded) and without (dashed line and grey shaded)
optical afterglow.  The distribution of bursts with afterglow includes
both \sax~and non \sax~bursts.  The vertical lines show the median
value of the column density for bursts with (solid line) and without
optical afterglow (dashed line).  The two distributions can belong to
the same parent population (at 94 per cent level, according to the KS
test).}
\label{fig:nh}}
\end{figure}

We have checked that local Galactic extinction does not play a crucial
role by comparing the hydrogen column densities in the direction of
detected afterglows with those in the direction of FOAs.
Figure~\ref{fig:nh} shows the comparison.  Applying a
Kolmogorov-Smirnov (KS) test (Press et al. 1992), we obtain that the
two distributions are drawn from the same parent population at the 94
per cent confidence level.

The visual inspection of Figure~\ref{fig:opt} reveals a clear
segregation of arrows from dots, the former being systematically
fainter than the latter at comparable times.  The eye impression can
be tested through several statistical tests. If we consider the whole
sample of OAs, a bidimensional KS test (Press et al. 1992) can be
applied.  The probability for the circles (empty $+$ filled) and the
arrows being derived from the same parent distribution is $P\sim$ 0.2
per cent ($\sim 3\sigma$). This test, however, may be biased due to
the systematic difference in the observing times between the OAs and
FOAs samples.  Since a correlation between the $R$ magnitude and the
detection time $t$ is present in the data, a monodimensional test can
be applied on the residual quantity:
\begin{equation}
M = R-\{A+B[log(t)-1.2]\},
\label{eq:relazio}
\end{equation}
where $A$ and $B$ are two parameters obtained by a linear fit of the
OA sample and the value 1.2 is the logarithmic average observing
time. Fitting Eq.~\ref{eq:relazio} to the \sax~OAs data, we obtain
$A=20.84\pm0.44$ and $B=2.86\pm1.2$. Using non-\sax~data only, we
find $A=19.84\pm0.54$ and $B=2.31\pm1.2$. Since the two results are in
agreement at the $1.5\sigma$ level, in the following we use the $A$
and $B$ parameters obtained in the cumulative fit of \sax~and
non-\sax~OAs. This yields:
\begin{equation}
A=20.3\pm0.3 \qquad\qquad
B=1.9\pm0.7
\label{eq:ab}
\end{equation} 
(see also Eq.~\ref{eq:rela}). It is however worth mentioning
that, should the earliest photometric point of GRB970508 be used
instead of the brightest one (see Tab.~\ref{tab:uno}), the difference
of the two subsamples would be increased to $\sim2.3\sigma$.

We obtain the following results. The distribution of the $M$ residuals
(Eq.~\ref{eq:relazio} with parameters from Eq.~\ref{eq:ab}) for the
\sax~and non-\sax~OAs is different from that of the FOAs at the 99.4 
per cent ($\sim 2.7\sigma$) level. If we consider, instead, the
distribution of the $M$ residuals for \sax~OAs and FOAs only, these
two differ at the 95 per cent ($\sim 2\sigma$) level.

Alternatively, one can fix the value of the parameter $B$ to the
median decay slope of GRB afterglows
($B=2.5\langle\delta_R\rangle=3.25$) and derive $A=20.3\pm0.3$. Using
these values of $A$ and $B$, the two tests described above yield
$P=99.5$ and $P=94$ per cent, respectively.  These numbers are very
similar to those obtained with the fitted slope $B$.

All the statistical tests discussed above do not assume any particular
shape for the probability distribution of OAs or FOAs but depend
on the assumption that \sax~OAs and non-\sax~OAs are drawn from the
same parent population. A more stringent test can be applied if we
assume that the residual quantity $M$ is gaussianly distributed around
the mean value 0 for \sax~OAs. We find that 4 FOAs out of 19 deviates
more than $2\, \sigma$. The probability for this is $P\sim 0.08$ per
cent ($\sim 3.3\sigma$).

This result shows that in many cases we failed to detect the optical
afterglow not because the search was conducted without the necessary
depth, but instead because the FOAs are indeed fainter than the
detected ones.  Yet, it is possible that FOAs are optically fainter
because intrinsically less energetic at all wavelengths, or because
they are more distant.  In order to check this, we compared the X-ray
and $R$ band flux densities of bursts with and without optical
detection 12~hours after the burst event.  For bursts with optical
detection, the magnitude at $t=12$~h has been computed taking into
account the measured flux decay with time while for FOAs an average
$\delta_R=1$\footnote{The parameter $\delta$ is defined through
$F_\nu(t) \propto t^{-\delta}$.} has been used.

\begin{figure}
\centerline{\psfig{file=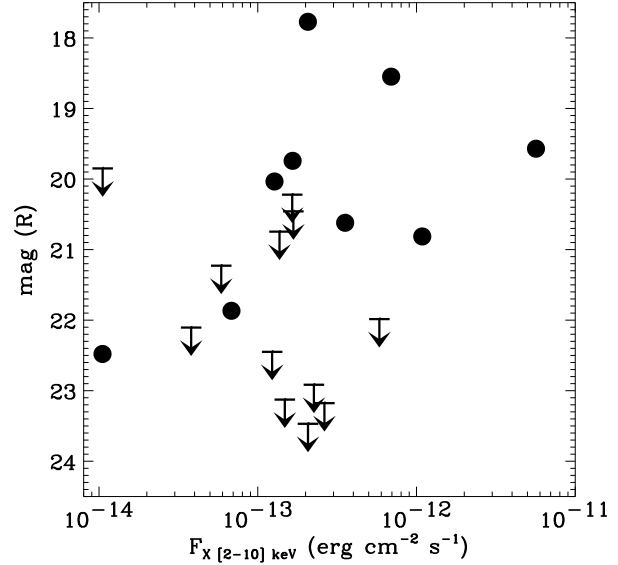,width=8cm}}
\caption
{{Optical $R$ band magnitudes vs X-ray flux for \sax~afterglows.
Dots are afterglows with both optical and X-ray detections while
arrows are upper limits for afterglows with X-ray detection but
without any optical detection.}
\label{fig:fxfo}}
\end{figure}

The result is shown in Fig.~\ref{fig:fxfo}, where dots correspond to
detected afterglows while arrows indicate upper limits.  In this
figure, the number of points is smaller than in Fig.~\ref{fig:opt}
because only those bursts with \sax~NFI observations have been
plotted, for consistency in the X-ray flux.  We can see that the
X-ray fluxes of FOAs are not systematically fainter than the fluxes
of afterglows with optical detection, indicating that FOAs are indeed
optically poor and define a different population with respect to
optically detected afterglows.

Given this conclusion, we now explore the possibility that FOAs are
intrinsically similar to bursts with detected optical afterglows, but
suffer from dust extinction due to the propagation of their photons in
a molecular cloud, where the burst explosion took place.

We here estimate a lower limit to the extinction that may cause the
FOAs to go undetected.  There are two ways of doing this.  A first way
is to consider the ensambles of detections and upper limits, add a
constant magnitude shift to all upper limits and recompute the KS
probability until a maximum value is reached.  This will give the
average value of the extinction required to have an undetectable
afterglow.  This procedure gives an average absorption in the $R$ band
of $\langle A_R\rangle\sim2.0$.  Dereddening all FOAs by this amount
maximizes the probability that all bursts belong to the same parent
population ($P\sim$35 per cent).

To compute the fraction of FOAs with respect to all bursts we must use
an homogeneous dataset.  We then use only bursts detected by the WFC
of {\it Beppo}SAX, excluding GRB~980425 and X-ray transients.  We are
left with 31 bursts, 19 of which are FOAs, yielding a fraction of 60
per cent.  Therefore an average absorption of 2 magnitudes in the $R$
band is needed for more than half of the bursts.

A different method to constrain the required absorption for FOAs is to
estimate a burst-by-burst absorption by computing the lack of
brightness with respect to an ``average afterglow'' $R$-$t$ relation
(see Eq.~\ref{eq:relazio} and~\ref{eq:ab} above):
\begin{equation}
R = 17.991+1.936 \, \log t
\label{eq:rela}
\end{equation}
where $t$ is given in hours.  This fit is shown in
Figure~\ref{fig:opt} with a solid line.

The dotted line in Figure~\ref{fig:mix} shows the integrated
distribution of the required $A_R$ to bring each upper limit on the
magnitude of FOAs to the magnitude given by Equation 1.  X-ray
transients have not been considered in this sample.  Since it is not
possible to quantify the local absorption that affects the detected
bursts, we conservatively assume that their local absorption is
negligible.  The fact that the dotted line saturates for $A_R \le 0.6$
is due to this assumption.  If the optical afterglows of the detected
bursts were locally absorbed, the dotted line would approach unity for
$A_R$ in the range shown in Fig. 4, making the discrepancy among the
distributions even larger and then making a more compelling case.

\section{Absorption in molecular clouds}

Absorption of several visual magnitudes in a molecular cloud is not
uncommon, and hence the hypothesis that failed optical detections are
due to absorption deserves a detailed analysis.

\subsection{Average cloud absorption}

As a first simple approach, we consider molecular clouds as uniform,
with standard dust to gas ratios.  A compilation of Galactic cloud
masses and sizes is given in Leisawitz et al. (1989), who analyze with
a systematic survey the CO emission around 34 young open clusters of
the Galaxy.  We have computed the column density
\begin{equation}
N_{\rm H}\, =\, {3\,M \over 4\pi\,m_{\rm p}\,a_{\rm maj}\,a_{\rm min}}
\label{eq:nh}
\end{equation}
where $M$ is the mass of the cloud and $a_{\rm maj}$ and $a_{\rm min}$
are the major and minor axis of the cloud.  For simplicity we have
assumed that all the matter is in pure atomic hydrogen.  This
assumption implies that the derived $N_{\rm H}$ is slightly
overestimated, but correct up to factors of order unity.

Analyzing the data of the same molecular clouds, Leisawitz (1990)
derives a maximum observed column density of molecular hydrogen $H_2$
(note that a factor of two difference with $N_{\rm H}$ is expected for
clouds with the same total mass) in each cloud.  Clouds were observed
through 3 to 144 lines of sight.  Since the interstellar medium (ISM)
in the clouds is not uniform, the maximum observed value is
significantly larger than the derived average one (see
Fig.~\ref{fig:mix}) and may be a better statistical indicator of the
column density where massive stars form.

\subsection{The Orion molecular cloud}

To better constrain the distribution of the expected extinctions
within a single star-forming molecular cloud we have analyzed the
observed extinction in O, B and A stars within the Orion molecular
cloud.  Data of $E_{B-V}$ have been taken from Lee (1968), who gives
the observed reddening for a sample of 196 stars.  This reddening has
been converted in absorption in the $V$ band ($A_V$) by adopting the
average shape parameter $R_V$ appropriate for the environment close to
hot, massive and young stars in the cloud (Lee 1968):
\begin{equation}
R_V \, \equiv\, {A_V \over E_{B-V} }\,  = \, 5.5
\end{equation}
from which $A_V = 5.5 \, E_{B-V}$ (see also Sect.~\ref{sec:compass}).

\section{Comparison with afterglow (failed) observations}
\label{sec:compass}

In order to compare the absorption properties of (and in) the
molecular clouds, as described above, with the $R$ band absorption
derived for FOAs, we must convert column densities and $V$ band
absorption in the $R$ band.

To convert column densities into dust absorption, we adopt the dust to
gas ratio given in Predehl \& Schmitt (1995):
\begin{equation}
A_V = { N_{\rm H} \over 1.79 \times10^{21}{\rm cm^{-2}} }
\end{equation}
To convert dust absorption values from a wavelength to a different
one, we use the analytic approximation for the dust extinction curve
given in Cardelli et al. (1989).

A final problem is represented by the fact that we do not know the
redshift of FOAs.  Therefore we do not know at which rest frame
wavelength we have to compute the extinction in the afterglow.  In
fact, what we observe in the $R$ filter effective wavelength
$\lambda_R$ has been emitted (and dust extincted) at a rest frame
wavelength $\lambda = \lambda_R/(1+z)$.  As a {\it zero-order}
assumption, we put all FOAs at a redshift $z=1$, close to the average
value of the detected afterglows.  In this case we obtain
$A_{R\,(z=1)} = 1.31\,A_V$.

\begin{figure}
\centerline{\psfig{file=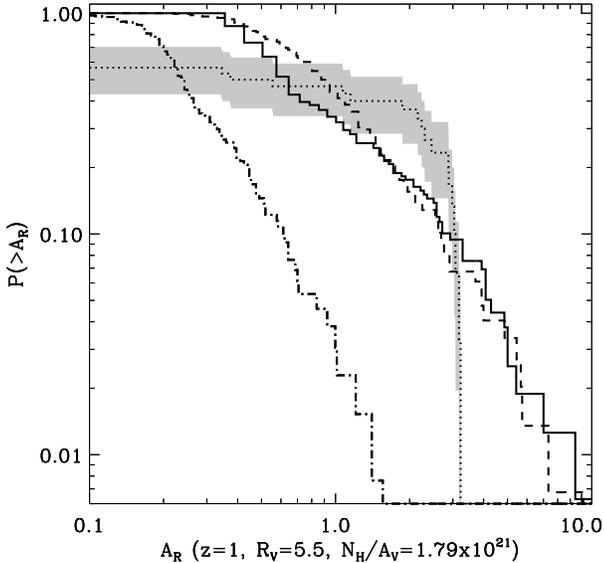,width=8cm}}
\caption{{Comparison between the estimated absorption in FOAs and
the absorption expected from molecular clouds.  Lines show the
integral distribution of the $R$ band absorption for the average
column density in molecular clouds (dot-dashed); for the peak column
density in molecular clouds (dashed) and for stars in Orion (solid).
See text for the conversion of $N_{\rm H}$ and $A_V$ in $z=1$ $R$ band
absorption.  The dotted line shows the integral distribution of the
lower limits of $A_R$ for FOAs, while the shaded area is the
$1\,\sigma$ confidence region.}
\label{fig:mix}}
\end{figure}

By adopting all the corrections described above, we can convert the
average $N_{\rm H}$, the peak $N_{\rm H_2}$ and the $A_V$ values of
molecular clouds into a distribution of expected $R$ band extinctions
for a burst at redshift $z=1$.  Figure~\ref{fig:mix} shows the result
of this conversion: the dot-dashed line shows the integral
distribution of $A_R$ values as derived from the average $N_{\rm H}$
of Galactic molecular clouds.  The dashed line shows the integral
distribution of $A_R$ as derived from the distribution of the peak
value of $N_{H_2}$ in the same clouds, while the solid line shows the
integral distribution derived from $A_V$ absorption of hot stars in
the Orion molecular cloud.

In order to compare the absorption required to obscure FOAs with the
distributions in Figure~\ref{fig:mix}, we consider first the average
$\langle A_R \rangle$ inferred for the bursts as a sample.

As detailed in Section 2 we need that 60 per cent of the lines of
sight to bursts are affected by 2 magnitudes of absorption.

Fig.~\ref{fig:mix} shows that if we use the average value of $N_{\rm
H}$ of the clouds, $A_R>2$ only in the $\sim$0.8 per cent of the
cases, while if we use the peak value of $N_{\rm H}$ this fraction
increases to $\sim$15 per cent.  We also have that $\sim$18 per cent
of the hot stars in Orion have $A_R>2$.  Note that all these values
refer to the absorption estimated in the observed $R$ band, but
assuming that the sources are at $z=1$.

These results imply that even if FOAs are located in the most absorbed
regions of molecular clouds, in several cases the corresponding
absorption is not enough to hide their optical afterglows.  Given our
assumptions, the statistical significance of this result is $\ge
3\,\sigma$: admittedly not extremely compelling, but consider that we
have used upper limits to measure the amount of $A_R$, which could
then be much larger.

Besides the integral distributions of $A_R$ observed in molecular
clouds, Fig.~\ref{fig:mix} shows (dotted line) the integral
distribution of the lower limits on $A_R$ derived for FOAs, with the
gray shaded area corresponding to the $1\,\sigma$ confidence region
for the same distribution.

The large negative deviations of the high and low absorption tails are
due to the intrinsic limitations of the lower limits sample, but the
difference around $A_R \sim 2$ with respect to the solid line is real,
even if significant at the $\ge 2\,\sigma$ level only.  This, again,
shows that the required absorption is larger than the absorption
associated on average to a molecular cloud at redshift $z=1$.

\section{Discussion}

By analyzing the properties of detected optical and X-ray afterglows
and the upper limits for failed detections, we show that the subset of
bursts without optical afterglow (FOAs) defines a different family.
This conclusion relies on several assumptions, like the homogeneity of
the \sax~and non~\sax~detected afterglows, imposed by the paucity of
the sample. Should some of these turn out to be wrong, the conclusion
would become statistically less stringent (see \S2 for further
details). We have investigated if this can be due to dust extinction
of optical radiation in a molecular cloud.  We find that this
hypothesis can only marginally account for the large fraction of FOAs,
and therefore we cannot exclude the possibility that FOAs are
intrinsically less luminous in the optical/UV band with respect to the
detected ones, and with respect to their own X-ray luminosity.

Consider also that we have been very conservative in our procedure,
because our results are based on considering {\it upper limits} on the
optical flux, and {\it peak} absorption columns expected in giant
molecular clouds.  The latter assumptions may well be too
conservative, if the dust is bound to evaporate when illuminated and
heated by the powerful optical/UV flash of the gamma-ray burst
(Waxman \& Draine 2000) and by its X-ray radiation (Fruchter et
al. 2000).  This dust sublimation is suggested for a sample of burst
afterglows (Vreeswijk et al. 1999c, Galama \& Wijers 2000), in which a
very large hydrogen column density $N_{\rm H} \gsim 10^{22}$
cm$^{-2}$, as estimated by X-ray data, is associated with almost no
optical extinction.  The results can be understood only in terms of a
dust to gas ratio $\sim 100$ times smaller than the Galactic average
value.  In turns, such low values of the dust to gas ratio can be
explained only if the dust has been completely sublimated in the
surroundings of the burst.  Indeed the theoretical models mentioned
above predict that dust can be destroyed by the burst emission out to
a radius comparable to the dimension of a typical molecular cloud (up
to a few tens of parsecs).  If this is the case, the material
responsible for absorption in FOAs is not the overdense cocoon
surrounding the star forming region, but the cloud as a whole (or even
less), and the discrepancy between the observed and measured value
(see dash-dotted line in Fig.~\ref{fig:mix}) becomes extremely
compelling.

An interesting way to assess whether dust is playing any role in FOAs
is to perform near infrared (NIR) follow-up of their $\gamma$ or
X-ray error boxes.  For instance, in the $K$ filter, absorption is
greatly reduced, so that only a very small fraction (less than 10 per
cent) of afterglows should show more than 1 magnitude of absorption,
in any of the adopted cloud models.  This is therefore a crucial test
to understand whether FOAs are due to dust absorption (less severe in
the near infrared) or to an intrinsic difference in the emitted
spectrum (that should be more severe in the NIR).  Some FOAs have been
indeed looked for in the NIR band, but the observations are still very
sparse and we lack any statistics to draw any meaningful conclusion.

NIR observations are thus strongly recommended as the key test for the
dust extinction hypothesis, especially after the launch of HETE II,
which will rapidly distribute accurate enough locations of bursts to
be promptly followed by ground based telescopes.  A more homogeneous
dataset, though, will have to await the launch of the Swift satellite,
foreseen in 2003.  The systematic follow-up with the on-board
optical telescope will provide a multiband spectroscopic database of
the first hours of optical afterglows.  Data of even higher quality
could be achieved if IR robotic telescopes (such as the one proposed
by the consortium of Brera, Rome and Catania Observatories, called
REM, for Rapid Eye Mount), will be in operation to complement Swift
observations from the ground.

A possibility to increase the absorption in the observed $R$ band
without invoking particularly dense molecular clouds is by allowing
for a higher redshift of the bursts.  This would make the afterglow
undetectable, especially if the redshift of the burst is particularly
high ($z\gsim 4$), so that the redshifted Lyman $\alpha$ break falls
in the $R$ filter.  In this case, again, near infrared ($JHK$)
observations should be unaffected by absorption and the optical
transient easily detectable. Estimates of the fraction of high
redshift GRBs (see, e.g., Porciani \& Madau 2001) predict however a
very small fraction of bursts (up to few per cent) at $z>3$, if the
GRB and star-formation rate are related. A more interesting way-out
is that the property of clouds at high redshift are different from
those of our Galaxy (see Ramirez-Ruiz Trentham \& Blain 2001), or
that the dust extinction curve changes its shape with redshift.

In conclusion, we have found that dust absorption due to a cloud with
properties similar to Galactic clouds is not a completely satisfying
explanation for bursts without a detected optical afterglow, and we
cannot rule out the possibility that they are due to an intrinsic
larger dispersion of optical fluxes with respect to the dispersion of
the X-ray fluxes (see also B\"oer \& Gendre 2000).

This, in turn, opens some exciting observational perspectives aiming
to disclose the nature of the burst progenitor: if bursts are indeed
associated with the final stages of stellar evolution and a
supernova-like event is associated to all bursts, then the search for
supernova signatures should be easier for bursts with an optical faint
afterglow, for which the SN lightcurve would not be polluted by the
flux of the afterglow.  If we assume SN1998bw (Galama et al. 1998c) as
a template supernova lightcurve, the expected magnitudes at maximum
should be roughly $I=24$ and $R=25$, easily detectable with a signal
to noise ratio of $\sim 10$ with an exposure time of only $\sim
10$~min with an 8 meter class telescope.

\section*{Acknowledgments}
We thank E. Feigelson, L. Stella and P. Saracco for useful discussions
and D. Malesani for help in preparing Table~\ref{tab:uno}.  We thank
Jochen Greiner for maintaining his web gamma-ray bursts archive.

\end{document}